# Predictive Optimized Model on Money Markets Instruments With Capital Market and Bank Rates Ratio

Bilal Hungund, NMIMS University, India*
Shilpa Rastogi, NMIMS University, India

## ABSTRACT

The money market and the capital market of the Indian financial markets have a symbiotic relationship in the development of the Indian economy. The nature and the characteristics of the markets differ to a large extent as the money market ensures liquidity in the system through the monetary policy by the regulators; capital markets propel and act as the engine driver for the economy in the long term. Therefore, the final throughput of the economy is the aggregation of the output of both the markets. Does that imply that the development of both markets is parallel in nature or is any one superior to the other or are they competitors? To understand the influence of one over the other the research was undertaken through a correlation matrix and time series model. A predictive model was further constructed for predicting the volume of money market instrument on the basis of fourteen days historical.

## 1. INTRODUCTION

As per the business standards, money market achieve tremendous growth in Indian financial markets just after the globalization initiative in the year 1991. Due to the advantages like high liquidity, borrowing or lending short term funds for a short time or a long time, diversity in interest rates and dichotomy in nature, popularize the market to gain the trust of the investors. To invest in Money Market, it provides various instrument options, for example, call money market (CMM), treasury bills (T-Bills), certificate of deposits (CDs), commercial papers (CPs), and money market mutual funds (MMMF). Due to this diversification in the market, it facilitates economic growth in India for the past few years.

Money Market and Capital Market focuses on trading of currencies and used to determine the flow of financial assets in monetary areas. So, there might be chances of having an either positive correlation or negative correlation among them. This can also be applicable in finding the relationship between bank rate ratios and money market instruments. This will basically prove that one of the variables, are dependent on any of another variables. Bank rate ratios, such as, cash reserve ratio (CRR), policy repo rate (PRR), and reverse repo rate (RRR) holds a significant position for impacting on each sector of countries economy. Hence changes in these ratios will definitely impact the financial markets.

*Corresponding Author

The volume of money market instruments changes independently with respect to time, hence it can be observed that volume of data may contain some trends or anomalies. Therefore, to check their stationarity in the data, time series model can be built which will help to identify those trends, and anomalies and to forecast for the future price.

In the end, merging these data with proper fortnight day transactions, a machine learning model is built which will predict the volume of money market instruments based on capital market value and bank rates ratio. All the statistics which will affect these variables will be interpreted.

## 2. LITERATURE REVIEW

There are numerous research studies that use similar indicators to forecast the direction of the stock market index. Much related work has been done on time series and modelling. A few reviews validated that machine learning and modelling have been the best technique to predict stock prices.

Ngabesong and McLauchlan (2019) "Implementing 'R' Programming for Time Series Analysis and Forecasting of Electricity Demand for Texas, USA" forecasted electric supply for Texas on the basis of historical data of one year on a one-point data from September 2016 to August 2017. The Auto Regressive Integrated Moving Average (ARIMA) model was used to estimate future predictions of electricity demand for Texas. It was concluded that the electricity demand would be on the rise for the next year and could also predict when peak shaving would be required.

Chauhan (2019), in his article on "Stock market forecasting using Time Series analysis" used the dataset consists of stock market data of Altaba Inc. which was retrieved from kaggle.com. from the year 1996 to 2017 for analysis. The Box Jenkins methodology (ARIMA model) was trained and predicted the stock prices on the test dataset.

Waqar et al (2017), "Prediction of Stock Market by Principal Component Analysis" conducted experiments on high dimensional spectral of 3 stock exchanges namely New York Stock Exchange, London Stock Exchange and Karachi Stock Exchange. The trend of three stock exchanges by using linear regression as a classification model and further to test the accuracy Principal component analysis, PCA was applied to predict the trend.

Roy et al. (2015), in their research paper "Stock Market Forecasting Using LASSO Linear Regression Model" proposed that the unique method of predicting financial market behaviour which was found to be far superior to the ridge linear regression model was through the Least Absolute Shrinkage and Selection Operator (LASSO) method based on a linear regression model. The model was experimented on the Goldman Sachs Group Inc. stock.

Mingyue and Yu (2016), in their research article "Predicting the Direction of Stock Market Index Movement Using an Optimized Artificial Neural Network Model" demonstrated that their model was most accurate to predict the direction of the next day's price of the Japanese stock market index with maximum accuracy by using the hybrid GA-ANN model. The authors applied two types of technical indicators to predict the direction of next day's Nikkei 225 index movement by adjusting the weights and biases of the ANN model using the GA algorithm and then tested the performance of the GA-ANN hybrid model by applying these two types of input variables and comparing the predictions with actual data.

Deng et al. (2011), "Combining technical analysis with sentiment analysis for stock price prediction" introduced a stock price prediction model. The stock price movements were modelled as a function of these input features and was solved as a regression problem in a Multiple Kernel Learning regression framework by them. The model extracted features from time series data and social networks for prediction of stock prices and Stock Market Forecasting using LASSO Linear Regression Model 373.

R.K. Dase and D.D. Pawar, (2010), "Application of Artificial Neural Network for stock market predictions: A review of literature," made a comprehensive review of neural network literature during the period 1991-2010. Neural Network has the ability to extract large sets of data. The authors concluded that Artificial Neural Network is highly efficient to predict stock index with traditional time series analysis.

Yoo et al. (2005), "Machine learning techniques and use of event information for stock market prediction: A survey and evaluation" in their work explored various global events and their concerns on forecasting stock markets. They found that integrating event information with the prediction model plays very significant roles for more exact prediction.

Abraham et al. (2005), "A.: Evolutionary multi objective optimization approach for evolving ensemble of intelligent paradigms for stock market modelling." experimented on NASDAQ stock market and S and P CNX nifty data to forecast stock prices. The authors proposed a genetic programming method for forecasting of stock market prices by combining two multi objective optimization algorithms with techniques like SVM and neural networks to get the best results.

Schumann and Lohrbach, (1993), "Comparing artificial neural networks with statistical methods within the field of stock market prediction." Considered both approaches stated in the title to predict short term predictions of the stock market. They further applied two models ARIMA and ANN for the prediction of the stock price which drew the conclusion that the results were, most accurate using modelling.

## 3. DELIVERABLES AND HYPOTHESIS

Based on the objective, three deliverables can be proposed.

### 3.1 Deliverable 1

To compare capital market instruments and bank rate ratios with money market instruments. This will identify that is there any relationship between money market and capital market or relationship between money market and bank rates. Hence, two hypotheses can be generated, i.e.:

**Hypothesis 1:** There is no relationship between money market instrument with capital market or bank rate ratios.
**Hypothesis 2:** There is a relationship between money market instrument with capital market or bank rate ratios.

Apart from the relationship, which type of relation the two variable holds, whether positive or negative.

### 3.2 Deliverable 2

To make time series analysis on money market. This deliverable also includes to find the best model which will forecast the volume of any money market instruments. To forecast the volume, auto regressive integrated moving average (ARIMA) model and Holt's winter model is implemented.

### 3.3 Deliverable 3

To build a machine learning model which will be predict the future price value of money market instruments, based on past ten years data with the help of principal component analysis (PCA), ordinary least square (OLS), gradient descent, regularization method (Ridge and Lasso), and sub-setting.

## 4. RESEARCH METHODOLOGY

To achieve all the objectives and deliverables certain standard steps are followed (Figure 1) which will result in building a better model and analysis.

Figure 1.
Flow Diagram to build and analyze the data model

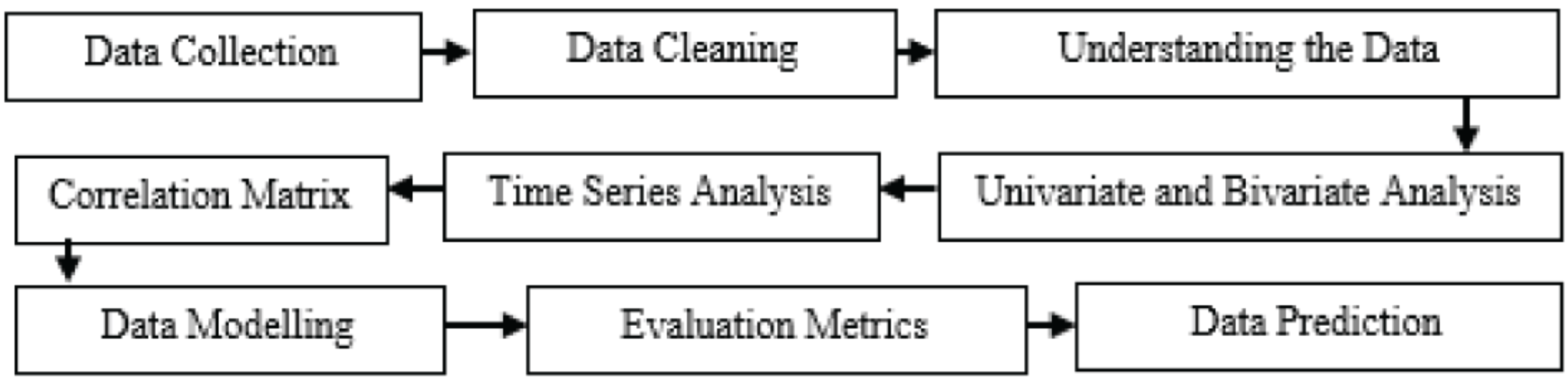

## 5. DATA ANALYSIS AND INTERPRETATION

### 5.1 Step 1: Data Collection

As per the requirement, four types of data are needed, any money market instruments data, stock price data (Sensex), bank rate ratios over a period of nine years (2011-2019), followed by including only fortnight day transactions.

Money market instruments and bank rate ratios data can be collected from RBI statistics website, stock data can be collected from any trusted and valid online business news websites.

All these data are collected in a raw file; hence cleaning process is needed (Figure 2).

### 5.2 Step 2: Data Cleaning

In this process, data are separated on the basis of dates i.e. segregating data in month, day, and year. This separation is required to make a time series analysis. In addition to this step, merging or joining

Figure 2.
Data Dictionary of required raw data

**Bank Rates Ratio**

| FortnightEnded | PRR | RRR | CRR |
|---|---|---|---|
| Sep. 13, 2019 | 5.4 | 5.15 | 4.0 |
| Sep. 6, 2019 | 5.4 | 5.15 | 4.0 |
| Aug. 30, 2019 | 5.4 | 5.15 | 4.0 |
| Aug. 23, 2019 | 5.4 | 5.15 | 4.0 |
| Aug. 16, 2019 | 5.4 | 5.15 | 4.0 |

**Stock Market Price**

| Month | Open | High | Low | Close |
|---|---|---|---|---|
| 2011-01-01 | 20621.61 | 20664.80 | 18038.48 | 18327.76 |
| 2011-02-01 | 18425.18 | 18690.97 | 17295.62 | 17823.40 |
| 2011-03-01 | 17982.28 | 19575.16 | 17792.17 | 19445.22 |
| 2011-04-01 | 19463.11 | 19811.14 | 18976.19 | 19135.96 |
| 2011-05-01 | 19224.05 | 19253.87 | 17786.13 | 18503.28 |

**Money Market Instruments**

| FortnightEnded | TotalAmountOutstanding | Volume | MinROI | MaxROI |
|---|---|---|---|---|
| 2019-08-31 | 497176.75 | 77603.30 | 5.37 | 13.39 |
| 2019-08-15 | 516900.55 | 123734.95 | 5.38 | 13.99 |
| 2019-07-31 | 509412.80 | 119806.65 | 5.69 | 14.47 |
| 2019-07-15 | 544981.40 | 81216.10 | 5.81 | 13.19 |
| 2019-06-30 | 503943.30 | 107691.05 | 5.79 | 11.89 |

of all the raw data will be done on the basis of *FortnightEnded* variable, as money market will only work on fortnight days. So, to join the tables, *FortnightEnded* will be the key. These merging will delete all duplicate values to overcome the redundancy problem (Figure 3).

## 5.3 Step 3: Understanding the Data

In this step, all the statistics will be measured to determine the spread ness in the data. Description of data includes determining mean, median, quartiles, standard deviation, maximum and minimum value of the data.

From Figure 4, it can be observed that, volume of CDs reaches to zero at some point, so this can be further analyzed that on which day it is zero, and what is the stock price and bank rates ratios.

## 5.4 Step 4: Univariate, Bivariate and Time Series Analysis

In this step, all different types of analysis are covered, results in different graphs to observe the money market instruments data.

After cleaning and merging the data twelve variables are considered, which are, fortnight day, volume of money market instrument, minimum and maximum return of investment (ROI) in money market on that particular fortnight day, open and close price of stock on that day, PRR, CRR, and RRR of that day. In this data, it can be inferred that, volume is the dependent variable and all other are independent variables.

In a univariate analysis, it can be observed that min ROI, Open and Close Price, PRR, and CRR data are normalized that is, no outliers are presents. This infers that, data are independently linear in nature. Followed by max ROI and CRR are not normalized that is there are some outliers present in the data for instruments.

In Figure 5, univariate analysis on max ROI of Commercial Paper is shown. Here it can be observed that the data follows a left-skewed distribution with some outliers which can be seen in the boxplot.

In a bivariate analysis, all independent variables will be compared with dependent variables to observe the differentiation between them. In the analysis it can be observe that volume is correlating with each and every other variable, which represents that, as one of the variable increases, other variable decreases, except bank rate ratio data. Comparing volume with bank ratio, it depicts that these variables are consistent on each other.

**Figure 3.**
**Data after final merging**

| FortnightEnded_y | Volume | MinROI | MaxROI | Year | Month | Day | Open | Close | PRR | RRR | CRR |
|---|---|---|---|---|---|---|---|---|---|---|---|
| 2019-08-16 | 16349.4598 | 5.4273 | 7.1940 | 2019 | 8 | 16 | 37387.18 | 37332.79 | 5.40 | 5.15 | 4.0 |
| 2019-08-02 | 13107.1423 | 5.6594 | 7.0873 | 2019 | 8 | 2 | 37387.18 | 37332.79 | 5.75 | 5.50 | 4.0 |

**Figure 4.**
**Statistical analysis on CDs data including stock price and ratio**

| | Volume | MinROI | MaxROI | Open | Close | PPR | RRR | CRR |
|---|---|---|---|---|---|---|---|---|
| **count** | 219.000000 | 219.000000 | 219.000000 | 219.000000 | 219.000000 | 219.000000 | 219.000000 | 219.000000 |
| **mean** | 25457.860592 | 7.584247 | 8.704856 | 26168.634247 | 26308.772283 | 7.098858 | 6.373973 | 4.284247 |
| **std** | 20228.913484 | 1.046139 | 1.278478 | 6841.888918 | 6892.878278 | 0.836680 | 0.568805 | 0.628520 |
| **min** | 0.000000 | 5.427300 | 6.490000 | 15534.670000 | 15454.920000 | 5.400000 | 5.150000 | 4.000000 |
| **25%** | 11123.122050 | 6.650000 | 7.750000 | 19452.050000 | 19395.810000 | 6.250000 | 5.750000 | 4.000000 |
| **50%** | 19296.108700 | 7.550000 | 8.700000 | 26681.470000 | 26638.110000 | 7.250000 | 6.250000 | 4.000000 |
| **75%** | 32067.608750 | 8.405000 | 9.780000 | 31156.040000 | 31283.720000 | 8.000000 | 7.000000 | 4.000000 |
| **max** | 101450.000000 | 9.640000 | 12.000000 | 39806.860000 | 39714.200000 | 8.500000 | 7.500000 | 6.000000 |

**Figure 5.**
**Univariate Analysis on Max ROI of Commercial Paper**

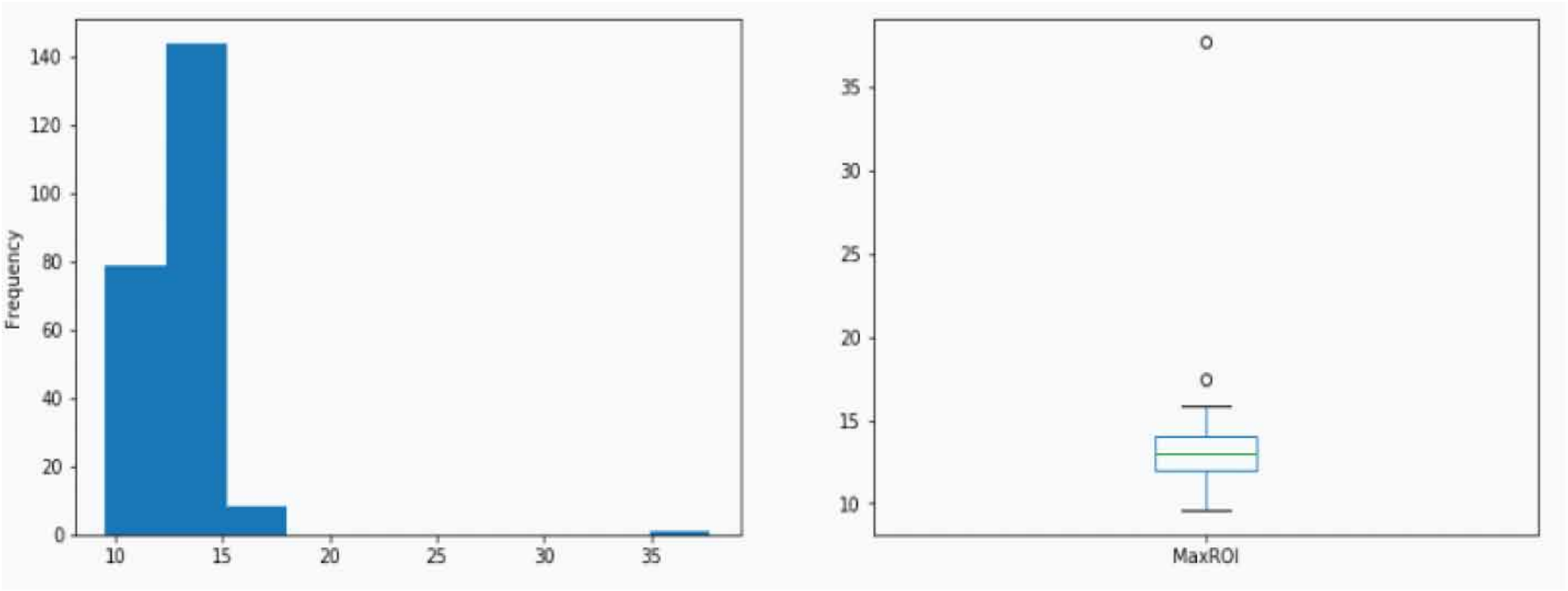

In Figure 6, open and close price of stock are compared with volume of certificate of deposits, which shows that as price (open or close) of stock increases, volume of CDs decreases.

Figure 7 shows that the ratios are probably consistent with volumes of money market instruments (including all instruments).

In this step basically all the analysis part is covered to determine the value of each data variables.

## 5.5 Step 5: Time Series Analysis

The time series analysis model is fit on the commercial papers data; hence all the analysis would imply a result with respect to commercial papers. The volume of commercial papers, i.e., a series, is a multiplicative time series because its monthly returns is not likely to be a normally distributed. This can be proved by analyzing Figure 8.

**Figure 6.**
**Bivariate on Stock Price and Volume of CDs**

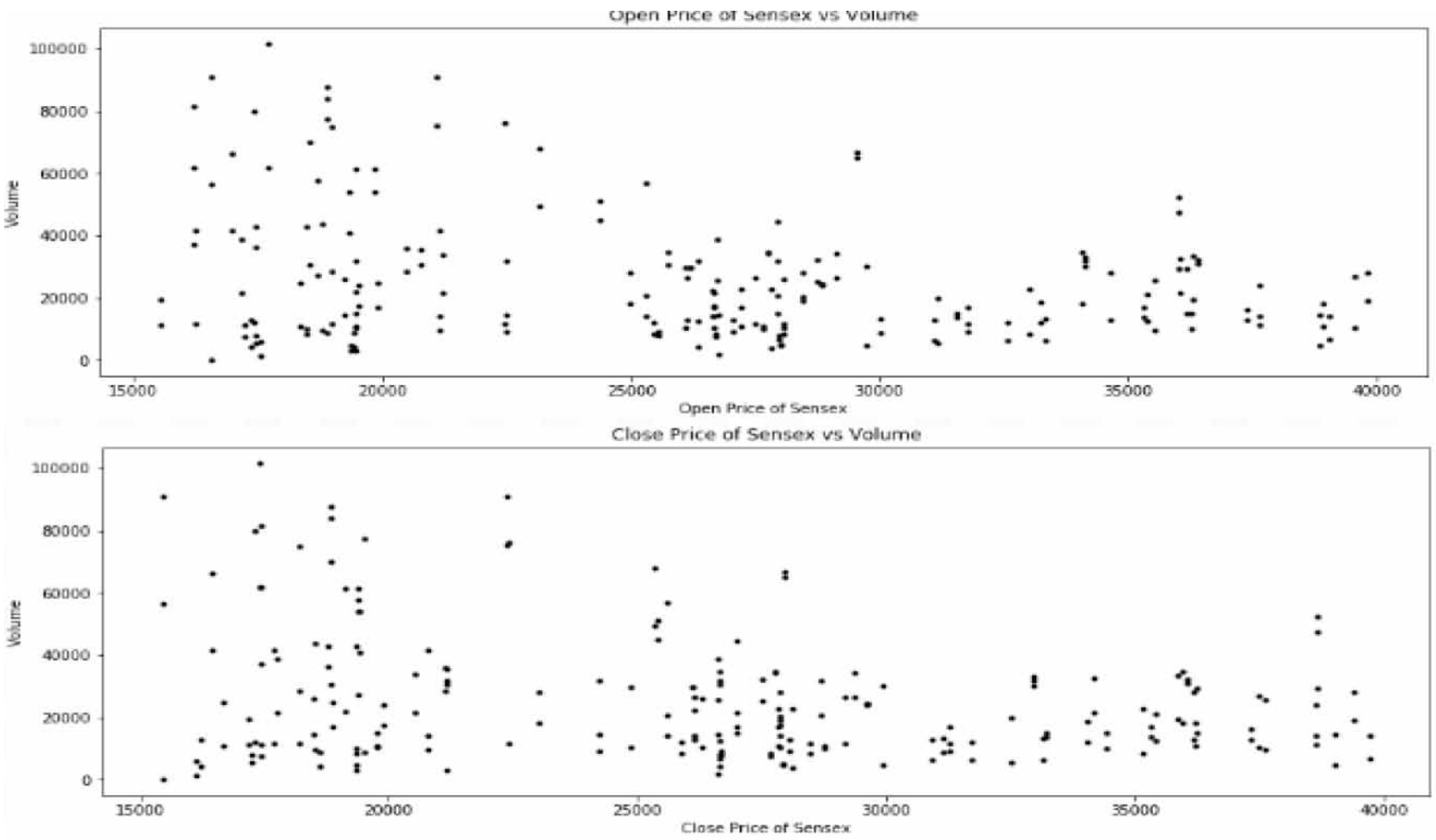

**Figure 7.**
**Bivariate on Repo-Rate and Volume of CDs**

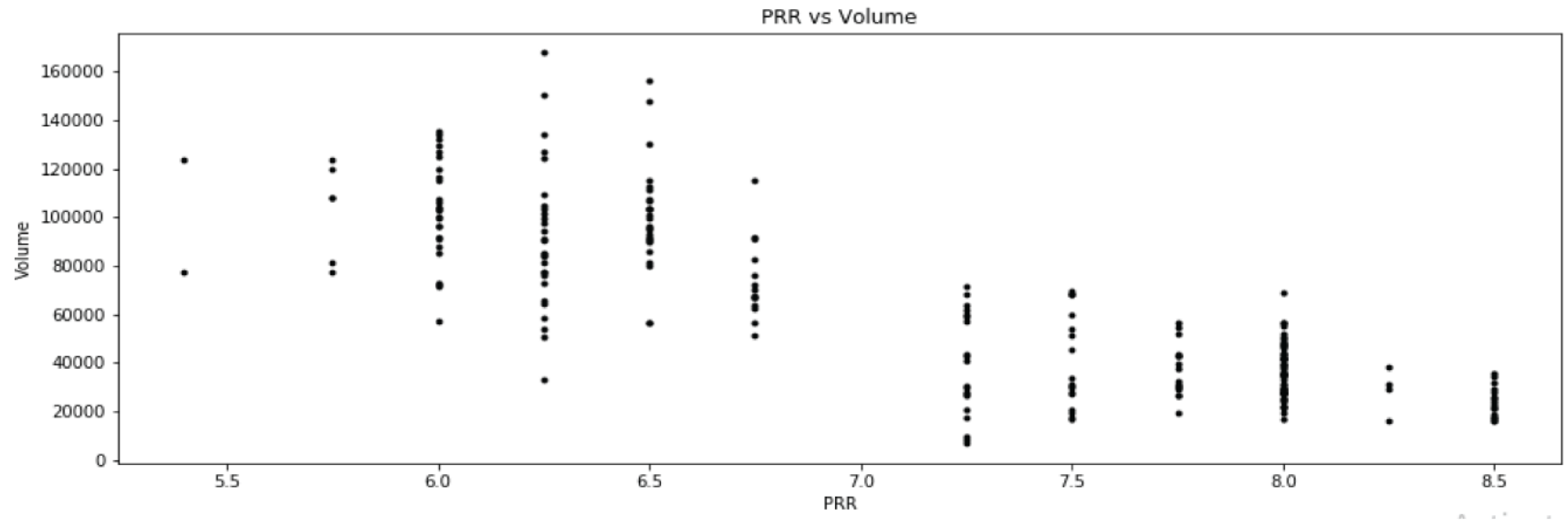

**Figure 8.**
**Histogram of commercial papers on monthly returns**

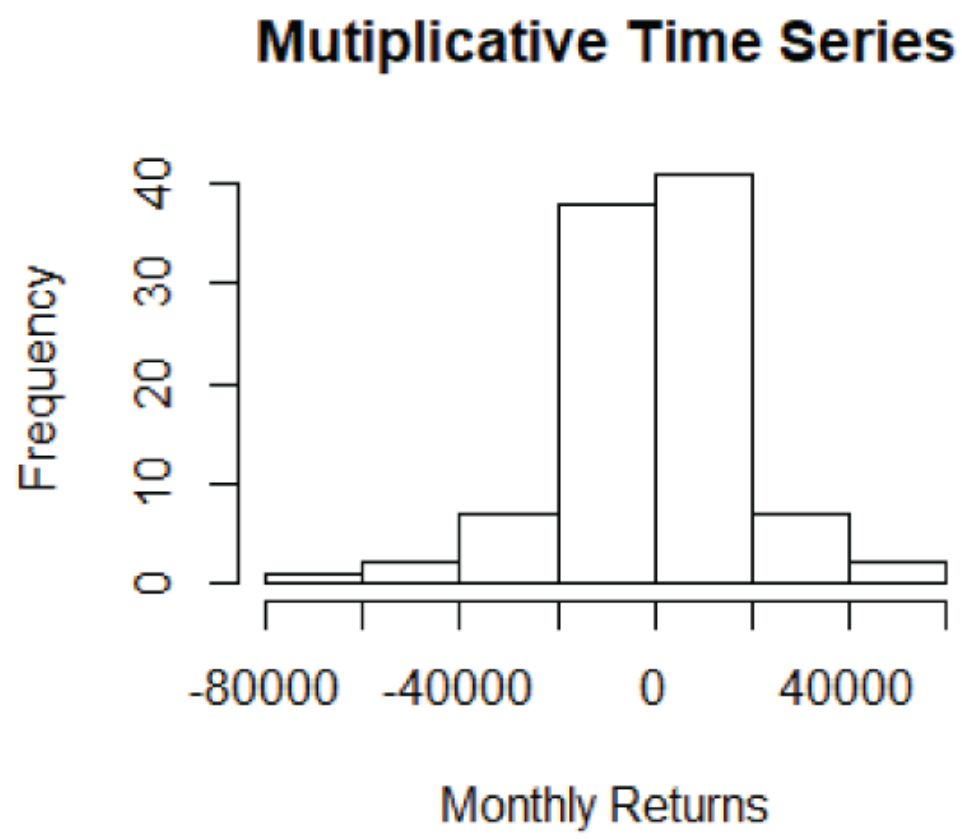

The series also contains a seasonality as well as an increasing trend for a past few years, Figure 9 will show the decomposition of the series to identify the seasonality and the trend.

It can be clearly observed that there is an increasing trend in the commercial papers for the past few years. To observe the above seasonality pattern Table 1 can be used to identify the pattern month-wise.

As all the pattern in the series are identified, next step is to check for the stationarity in the data using the Augmented Dickey Fuller Test, and after applying the test results shown in Table 2 are generated.

Table 2 shows that as p-value $< 0.05$ hence null hypothesis is rejected and the series is stationary, with lag order or AR (4).

Final step in time series analysis is to forecast the volume of commercial paper for next n months (n = 12). To perform the forecasting, two models are selected i.e. ARIMA model and Holt Winters model.

In an ARIMA model, to determine the best lag order of AR and MA, partial auto correlation function (PACF) is constructed.

From Figure 10, it shows that, the best lag order for AR or MA can be either 1 or 2 because all other lags are residing below the threshold line. To verify the above lag order or to get the best lag order it is various order of lags are tested in the ARMA model, which is preferably known as AUTO. ARIMA function. This function will give the best order of lags and minimal AIC and BIC value. The results in Table 3 are obtained from the AUTO.ARIMA.

**Figure 9.**
**Decomposition of series into trend, seasonality and error**

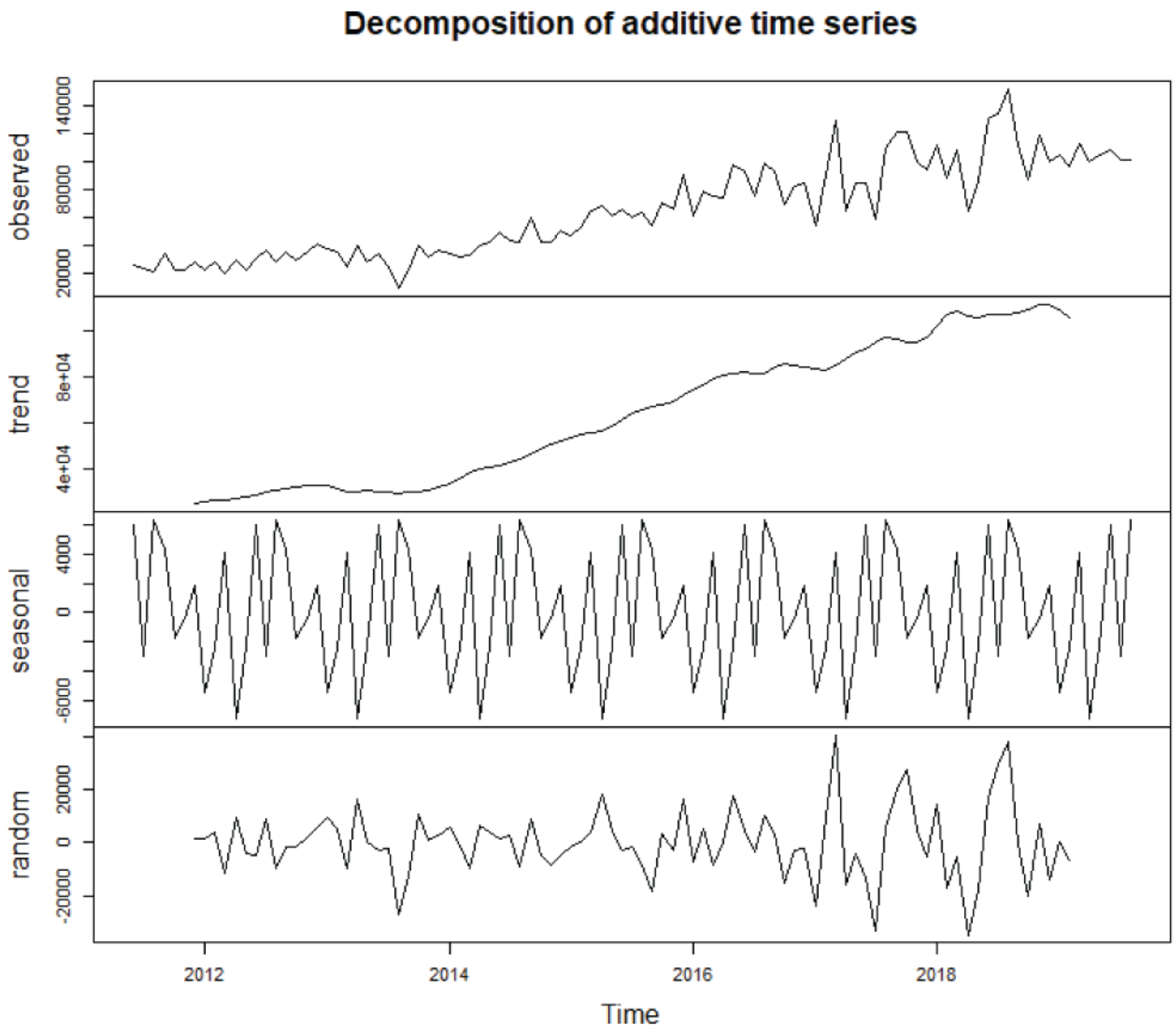

**Table 1.**
**Seasonality of the commercial papers**

| Month | Seasonality |
|---|---|
| January | -5519.32 |
| February | -2431.84 |
| March | 4148.07 |
| April | -7317.18 |
| May | -2282.32 |
| June | 5955.17 |
| July | -2945.59 |
| August | 6346.44 |
| September | 4361.82 |
| October | -1762.76 |
| November | -443.16 |
| December | 1890.69 |

**Table 2.**
**Results of ADF test on commercial papers**

| Parameter | Value |
|---|---|
| ADF Test Statistic | -3.554 |
| Lag order of AR | 4 |
| p-value | 0.049 |
| Null Hypothesis | Rejected, series is stationary |

**Figure 10.**
**PACF for best lag order of AR and MA**

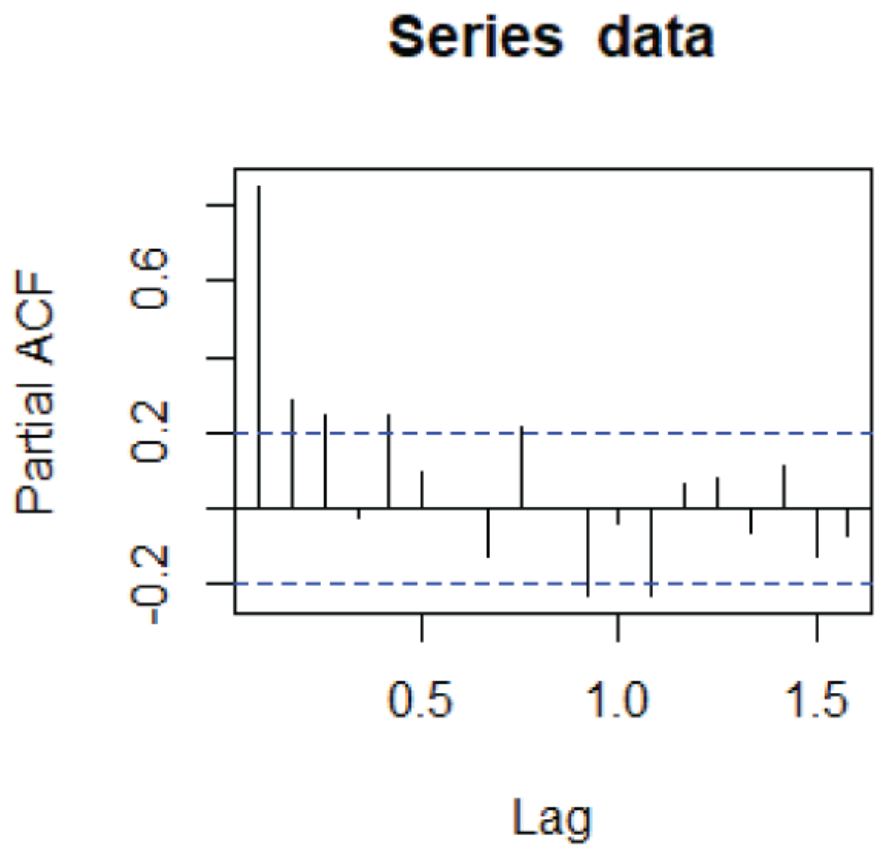

To further optimize the model, Holt Winters model with exponential smoothing is applied on the data to compare the results with ARIMA Model.

The models in Figure 11 are used to forecast the volume of commercial paper for the next 12 months, where red line denotes Holt Winters model and blue line ARIMA Model

After observing Figure 12, it can be concluded that Holt Winters Model is more reliable than ARIMA Model, as Holt Winters approach includes the error while predicting the value as well as exponential smoothing assigned a weight over time in a series.

**Table 3.**
**AUTO.ARIMA on commercial paper**

| ARIMA (1, 1, 2) with drift | | | | |
|---|---|---|---|---|
| **Parameter** | **AR-1** | **MA-1** | **MA-2** | **Drift** |
| Coefficients | -0.55 | -0.11 | -0.66 | 921.35 |
| Std. Error | 0.187 | 0.156 | 0.120 | 222.69 |
| | **AIC** | 2159.56 | **BIC** | 2172.49 |

**Figure 11.**
**Holt Winters applied on train data with exponential smoothing**

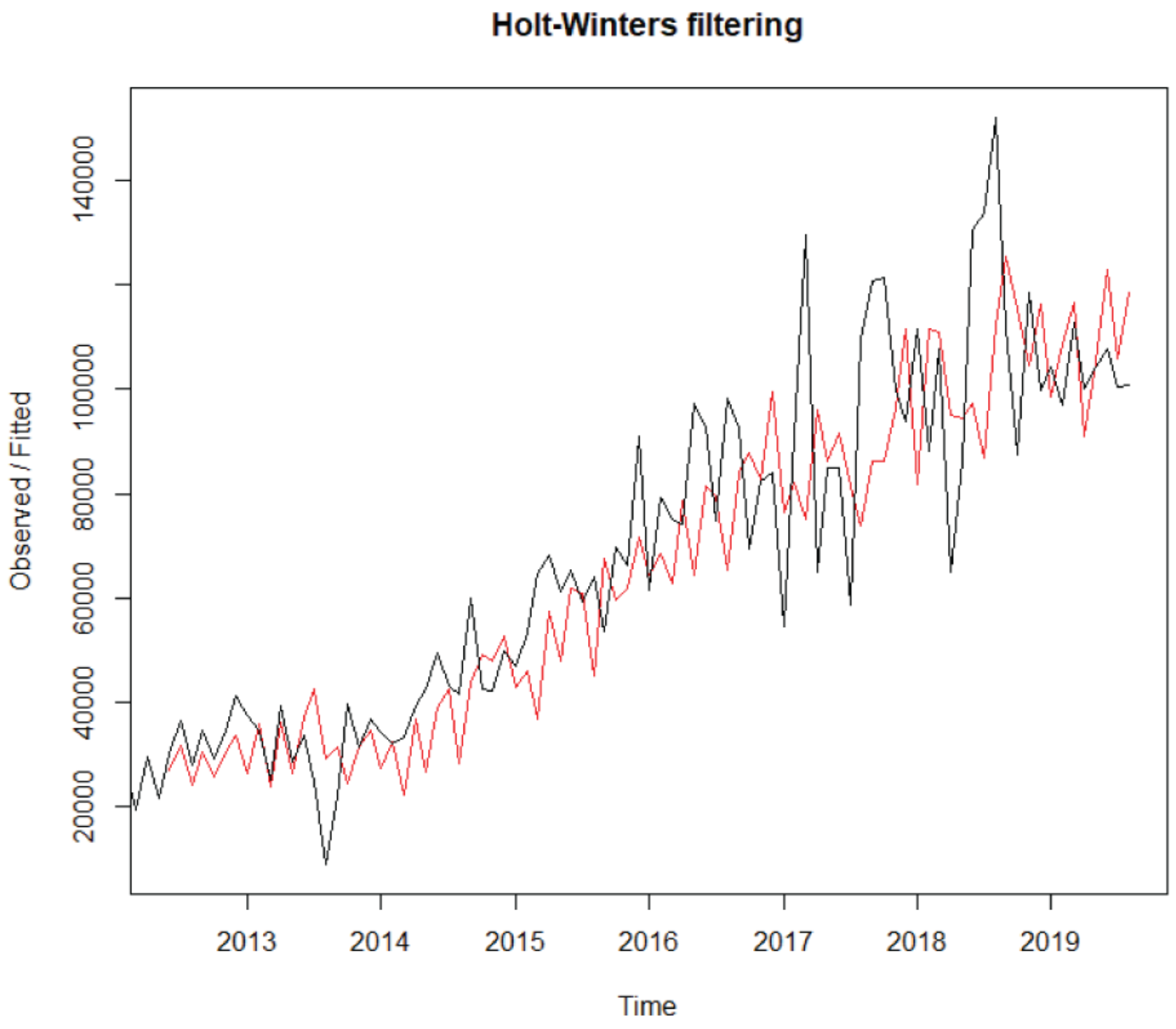

**Figure 12.**
**Prediction of commercial paper volume for the next 12 months**

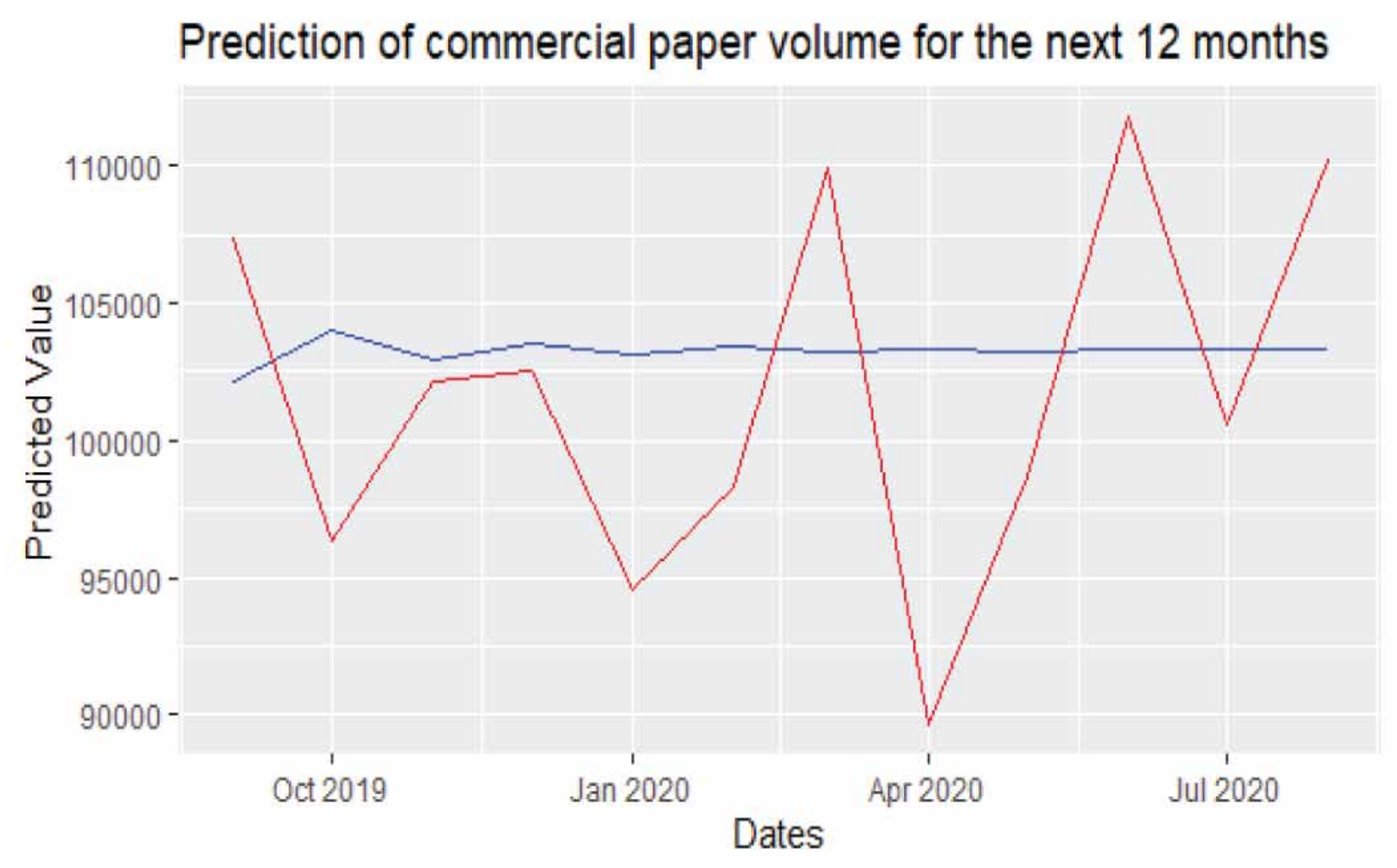

Therefore, to forecast the volume of money market instruments, Holt Winters model gives the best results and minimal error rates.

## 5.6 Step 6: Correlation Matrix

In this step, descriptive statistic is analyzed for dependent data against independent data.

In Figure 13, it can be observed that volume of commercial paper contains positive correlation with stock prices, that is, as volume increases stock price also increases or vice versa, whereas, volume contains negative correlation with bank rates ratio, that is, as volume increases bank rates ratio decreases or vice versa. Concluding same for certificate of deposit correlation matrix, here the difference is, volume contains negative correlation with stock prices and positive correlation with bank rates ratios.

## 5.7 Step 7: Data Modelling

After proper analysis of data, it can be concluded that combination of these data can help building a machine learning model, which will predict the future volume (dependent data) of any instruments based on past transactions. So, basic machine learning model can help to predict the goal.

From Figure 14. it can be observed that all variables are highly correlated among each other, this can lead to a problem of low variability in an axis, therefore to get high variability, PCA can be applied which will reduce the dimension, increase the model results, and overcome the model from overfitting.

In Figure 14, it can be clearly seen that, majority of variability lies in one direction, i.e. magnitude of a data is not changing, which means in both the axis it is difficult to find which quadrant of graph consists more variability. To showcase the change in magnitude of data, PCA is applied which results in high variability in one x-axis. This signifies that, as data are highly correlated than variability among the data is less, this would create a penalty error in evaluation of model, thus using PCA model it will be more tuned and high variability can be picturized, to increase the accuracy. PCA also requires standardization of data.

PCA in financial markets can be derived as:

**Figure 13.**
**Correlation Matrix**

Correlation Matrix of Commercial Paper

| | Volume | MinROI | MaxROI | Open | Close | PRR | RRR | CRR |
|---|---|---|---|---|---|---|---|---|
| Volume | 1.000000 | -0.795582 | -0.178074 | 0.849193 | 0.851875 | -0.814417 | -0.727022 | -0.479210 |
| MinROI | -0.795582 | 1.000000 | 0.146468 | -0.788679 | -0.785485 | 0.903864 | 0.843762 | 0.446812 |
| MaxROI | -0.178074 | 0.146468 | 1.000000 | -0.245760 | -0.258999 | 0.181386 | 0.197936 | 0.241595 |
| Open | 0.849193 | -0.788679 | -0.245760 | 1.000000 | 0.989235 | -0.832273 | -0.733223 | -0.567116 |
| Close | 0.851875 | -0.785485 | -0.258999 | 0.989235 | 1.000000 | -0.827590 | -0.724817 | -0.577689 |
| PRR | -0.814417 | 0.903864 | 0.181386 | -0.832273 | -0.827590 | 1.000000 | 0.960676 | 0.528925 |
| RRR | -0.727022 | 0.843762 | 0.197936 | -0.733223 | -0.724817 | 0.960676 | 1.000000 | 0.572418 |
| CRR | -0.479210 | 0.446812 | 0.241595 | -0.567116 | -0.577689 | 0.528925 | 0.572418 | 1.000000 |

Correlation Matrix of Certificate of Deposits

| | Volume | MinROI | MaxROI | Open | Close | PRR | RRR | CRR |
|---|---|---|---|---|---|---|---|---|
| Volume | 1.000000 | 0.360345 | 0.461124 | -0.283833 | -0.275479 | 0.331397 | 0.299537 | 0.180556 |
| MinROI | 0.360345 | 1.000000 | 0.878885 | -0.784829 | -0.776506 | 0.920927 | 0.869313 | 0.494652 |
| MaxROI | 0.461124 | 0.878885 | 1.000000 | -0.643462 | -0.640831 | 0.758889 | 0.733391 | 0.508839 |
| Open | -0.283833 | -0.784829 | -0.643462 | 1.000000 | 0.988109 | -0.806977 | -0.691260 | -0.563629 |
| Close | -0.275479 | -0.776506 | -0.640831 | 0.988109 | 1.000000 | -0.799534 | -0.679418 | -0.581244 |
| PRR | 0.331397 | 0.920927 | 0.758889 | -0.806977 | -0.799534 | 1.000000 | 0.954968 | 0.480057 |
| RRR | 0.299537 | 0.869313 | 0.733391 | -0.691260 | -0.679418 | 0.954968 | 1.000000 | 0.505640 |
| CRR | 0.180556 | 0.494652 | 0.508839 | -0.563629 | -0.581244 | 0.480057 | 0.505640 | 1.000000 |

**Figure 14.**
**Variability in CDs data**

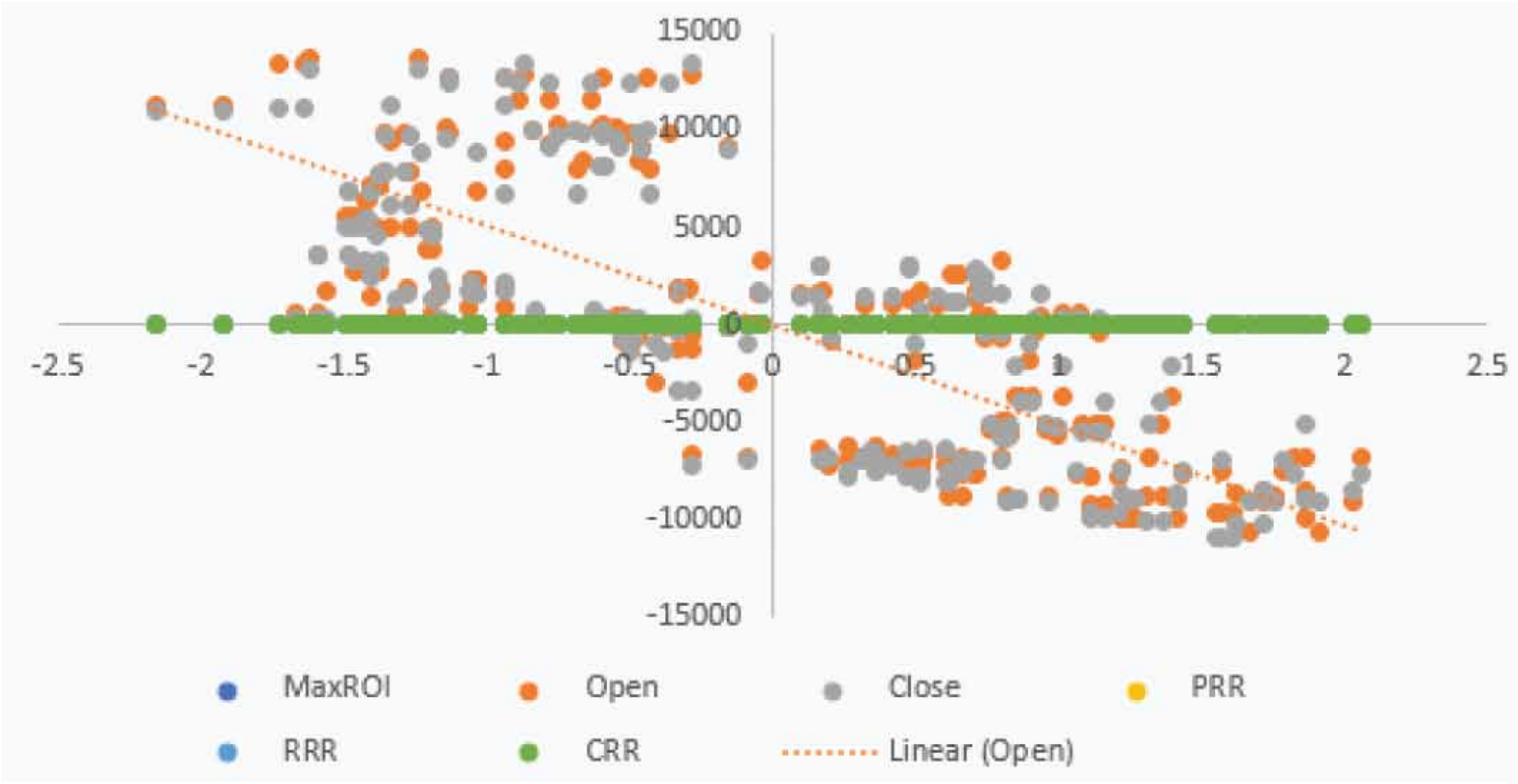

$$A * S_w = \lambda A$$

where, A is predictors, λ is eigen value of A, $S_w$ is Eigen Vector of A.

Through this equation eigen value and eigen vector can be computed for PCA, but before that covariance matrix is calculated by multiplying A and $A^T$.

Eigen values and variance ratios of PCA is shown in Table 4.

From Table 4, PCA concludes that Min. ROI (PCA 1) holds a majority of variability with 99% and this column can be further used to build a model with reduced dimension.

In Figure 15, it represents that how PCA 1 holds a majority of variability in projected line.

In financial terms, this infers that only Minimum ROI is necessary to build the model as this is the only factor which can be impacted on capital market and bank rates ratios. This can be due to that CDs are highly correlated with Sensex and Repo Rates.

**Table 4.**
**Market variations in different periods**

| PCA Components | Eigen Value | Variance Ratio (Cumm. value) |
|---|---|---|
| Min. ROI | 93762428.04 | 0.994054581 |
| Max. ROI | 560788.9875 | 0.005945397 |
| Open | 1.459401996 | 1.55E-08 |
| Close | 0.289052881 | 3.06E-09 |
| PRR | 0.232297229 | 2.46E-09 |
| CRR | 0.057817485 | 6.13E-10 |
| RRR | 0.010138413 | 1.07E-10 |
| **Total** | **94323219.07** | **1.00** |

**Figure 15.**
**PCA in CDs data**

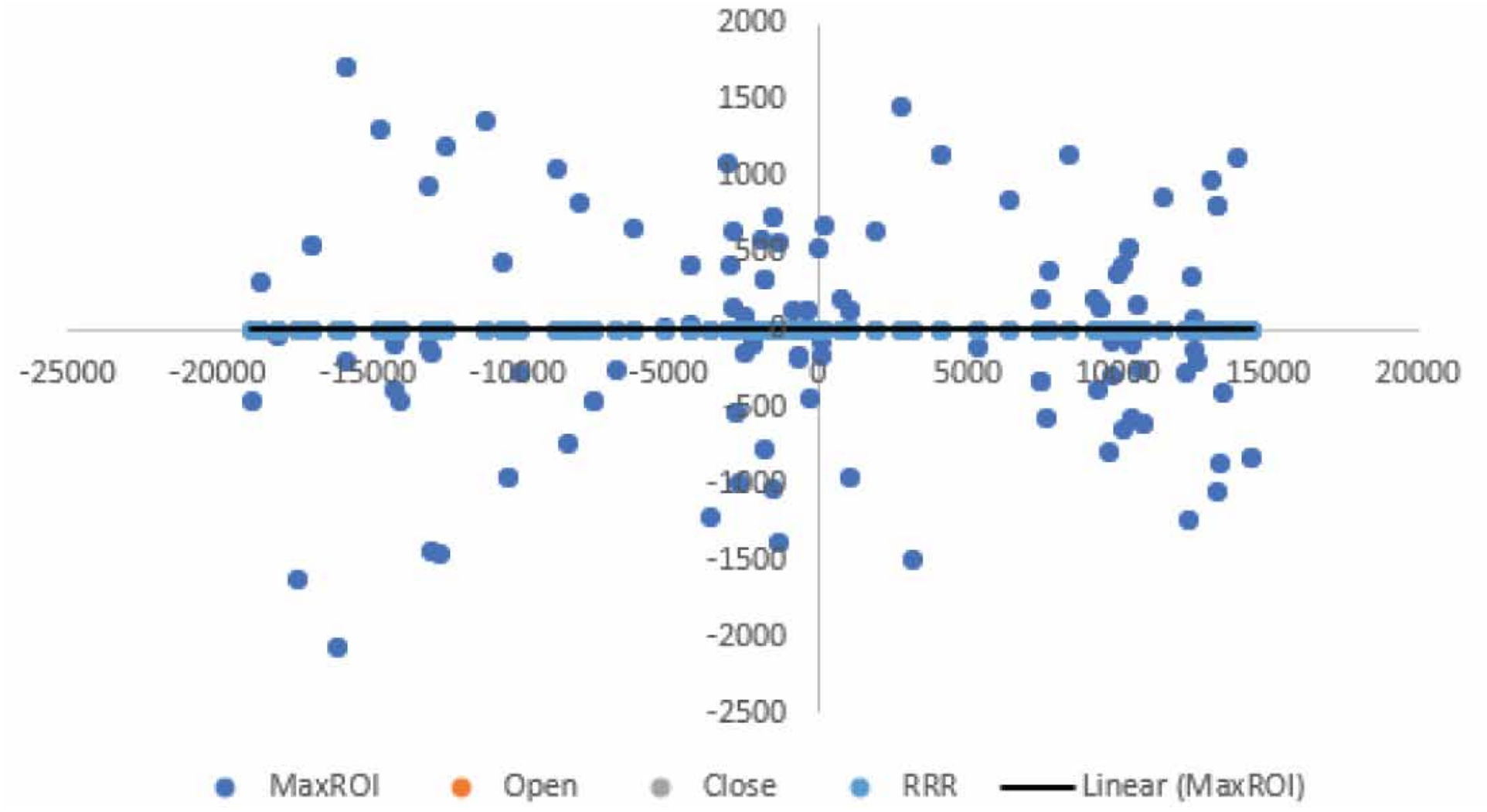

Hence with all these PCA Components data, model can be built which will definitely results in more accurate scores.

Before modelling of data, PCA components is extracted from the model. After extraction, segregation of data will be done, therefore dividing the data into two sets, train and validation part. To build the model, 80% of data are considered as training part, and 20% of data considered as validation part for evaluation of model.

### *5.7.1 First Model: Linear Regression Using Ordinary Least Square (OLS)*

In this method, simple linear regression model is performed using OLS approach, which results in Table 5.

From Table 5, it perceives that variable Max ROI of money market and close price of Sensex is not significant in the model as their p-value is greater than 0.05 which infers that their level of confidence in the model is less than 90% and therefore it should be ignored while building the optimized model so that accuracy would be minimized. Also, residuals of model are quite linear as the projected line is tending from low to high (Figure 16).

**Table 5.**
**Summary of OLS in CDs data**

| Variables | Coefficients | p-value | Confidence | Significant |
|---|---|---|---|---|
| (Intercept) | 2.567e+04 | < 2e-16 | 100% | Yes |
| Min. ROI | 6.031e-01 | 1.51e-05 | 100% | Yes |
| Max. ROI | 9.375e-01 | 0.5911 | - | No |
| Open | 5.217e+03 | 2.46e-06 | 100% | Yes |
| Close | -2.606e+03 | 0.2708 | - | No |
| PRR | -4.994e+03 | 0.0580 | 90% | Yes |
| RRR | -1.139e+04 | 0.0336 | 95% | Yes |
| CRR | -2.412e+04 | 0.0696 | 90% | Yes |

**Figure 16.**
**Residuals in Projected Line of Linear Regression**

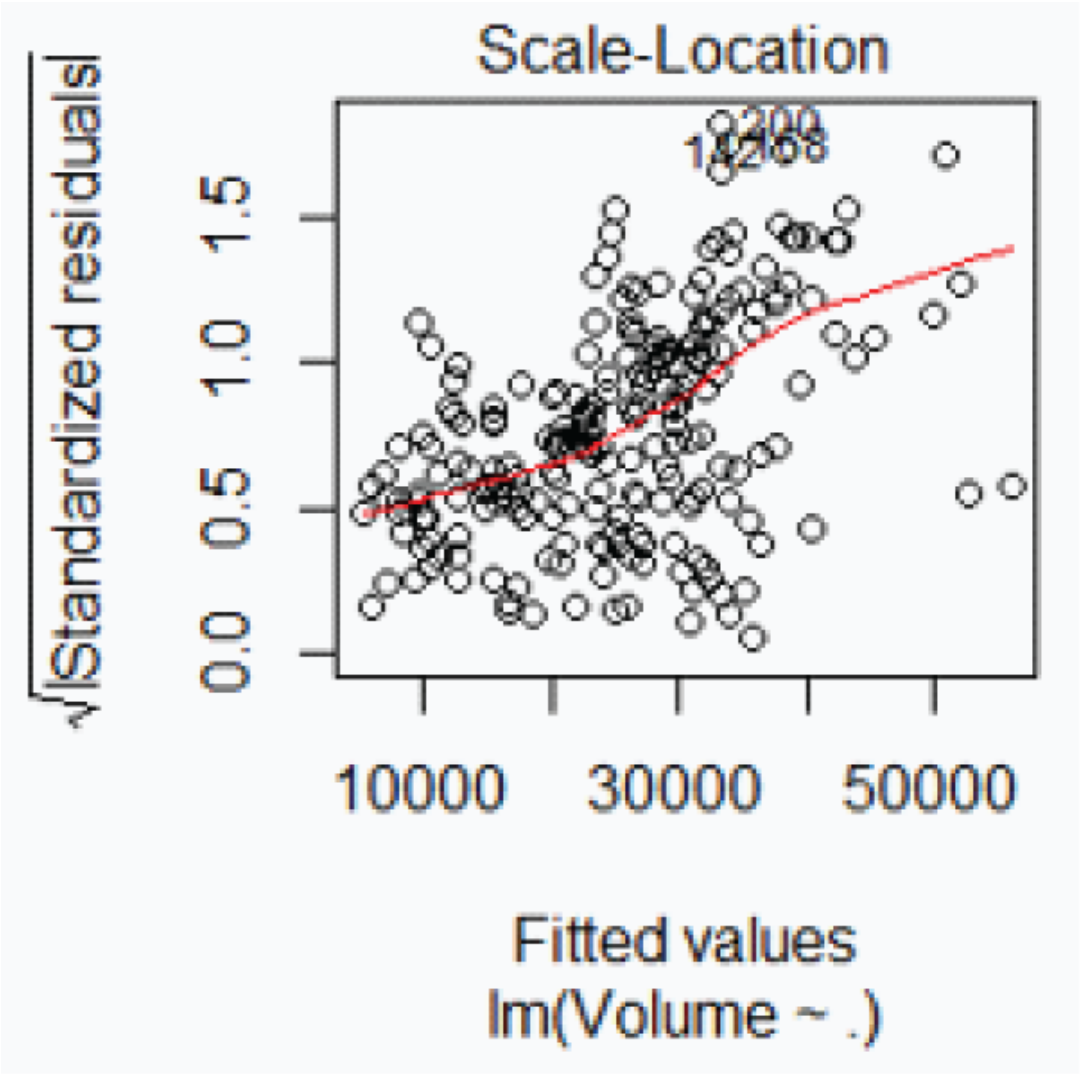

Since this is not an optimized regression model to predict the target variable as its RMSE (comparison in table number 11) is high as per the standard, and therefore gradient descent approach can be implemented for better results.

### *5.7.2 Second Model: Linear Regression Using Gradient Descent*

The main objective of gradient descent is to minimize the standard error, as the accuracy in OLS method is high. Gradient descent uses linear algebra to calculate the predictions with a factor known as learning rate. Learning rate helps the model to tune and make them more optimized, and because it uses tuning several iterations are computed and the best iteration of coefficients are considered to project the line.

Gradient Descent for calculating the intercept is given as, considering data in a matrix form, for Intercept:

$$\theta_0 := \theta_0 - \alpha * \frac{1}{m} * \left(H\left(\theta\right) - Y\right)$$

For the coefficients (predictors) it is given as:

$$\theta_i := \theta_i - \alpha * \frac{1}{m} * \left(H\left(\theta\right) - Y\right) * X_i$$

where H(θ) is given as:

$$H\left(\theta\right) = \theta_0 + \theta_i X_i$$

Therefore, to compute the cost function it is given as:

$$J(\theta) = \frac{1}{2m} * \left(H(\theta) - Y\right)^2$$

where, $X_i$ is predictors (in matrix), Y is target variable, m is number of observation and α is learning rate.

Through this approach minimal accuracy can be achieve compared with OLS method. Table 6 shows the constraints applied on a data which tends to good results.

Table 7 shows the thetas computed for the optimized model.

Considering all these variables which are not essential has impacted on the model and hence to reduce the complexity of the model, optimal variable selection method can be perform, this method can be achieve by implementing Regularization (L1-Norms) and Sub-setting.

To apply regularization, lasso regression can be performed which will shrink the coefficients towards zero by penalizing the linear regression (gradient descent) model with a penalty term as, L1-norms. Here, lasso regression is projected as:

$$RSS_L := RSS + \sum \lambda \left|B_i\right|$$

where, λ is penalty amount for tuning and $B_i$ is predictors.

Also, it is always suggested that, user-defined lambda should be considered rather than system generated lambda, this will help to adjust the amount of the coefficients shrinkage. To get better results 100 lambdas from 1e10 to 0.002 are generated for the data.

Figure 17 shows that many variables are tending towards zero for better accuracy, for some variables there is a drastic change in the coefficients. Before finalizing the model, cross validation will help to achieve minimal error rate and will compute the best lambda constant for the model. Here, CV initialize to five folds. Table 8 shows the results achieved in lasso regression.

**Table 6.**
**Constraint of gradient descent in model**

| Constraints | Value |
|---|---|
| Learning Rate λ | 0.02 |
| Max Iteration | 2000 |
| Scaling of Data | True |

**Table 7.**
**Coefficients of each variable**

| Variables | Coefficients |
|---|---|
| (Intercept) | 25439.99 |
| Min. ROI | 6750.06 |
| Max. ROI | 664.09 |
| Open | 6821.68 |
| Close | -2530.34 |
| PRR | -3069.22 |
| RRR | -3526.18 |
| CRR | -1020.17 |

**Figure 17.**
**Lasso Regression Coefficients Shrinkage**

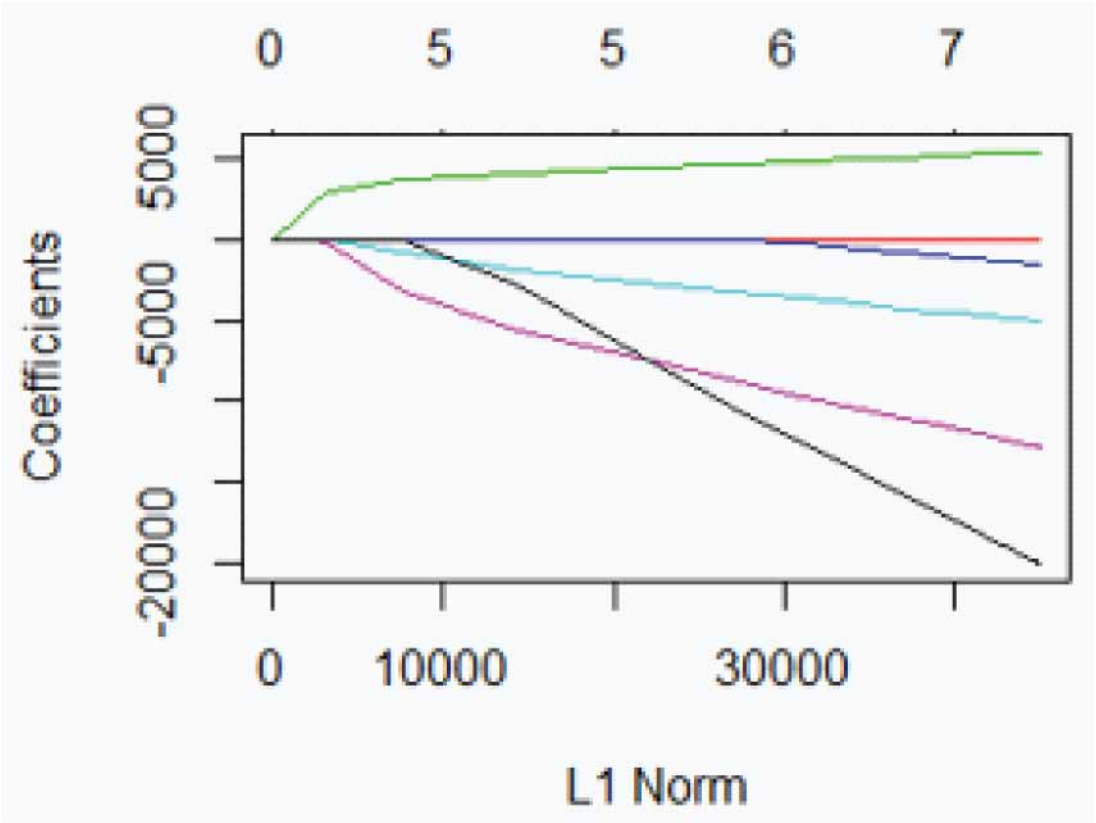

Coefficients of each variables after applying the Lasso Regression are shown in Table 9.

Table 9 shows that two variables i.e. Close and Maximum ROI are not impacting on model accuracy and so it can be neglected while building the model. Therefore, with these variables model is build and minimal accuracy is achieved comparing with normal linear regression approaches.

Figure 18 shows minimal accuracy impacted while applying cross validation on lasso regression, where the best lambda for shrinking is estimated.

Another method which can be used to reduce the accuracy and model complexity is sub-setting (also known as, feature selection). In this method, subset for every variable are created and the best subset of variables is captured which gives the minimal accuracy and optimized model.

**Table 8.**
**Results of lasso regression**

| Results | Value |
|---|---|
| Best Lambda λ | 1232.847 |
| No. of Variables Shrink | 2 |

**Table 9.**
**Coefficients of each variables after applying lasso regression**

| Variables | Coefficients |
|---|---|
| (Intercept) | 25762.4379802 |
| Min. ROI | 0.4658644 |
| Max. ROI | . |
| Open | 4424.3725265 |
| Close | . |
| PRR | -2626.2532256 |
| RRR | -7359.4979595 |
| CRR | -7077.8359284 |

**Figure 18.**
**Cross Validation on Lasso Regression**

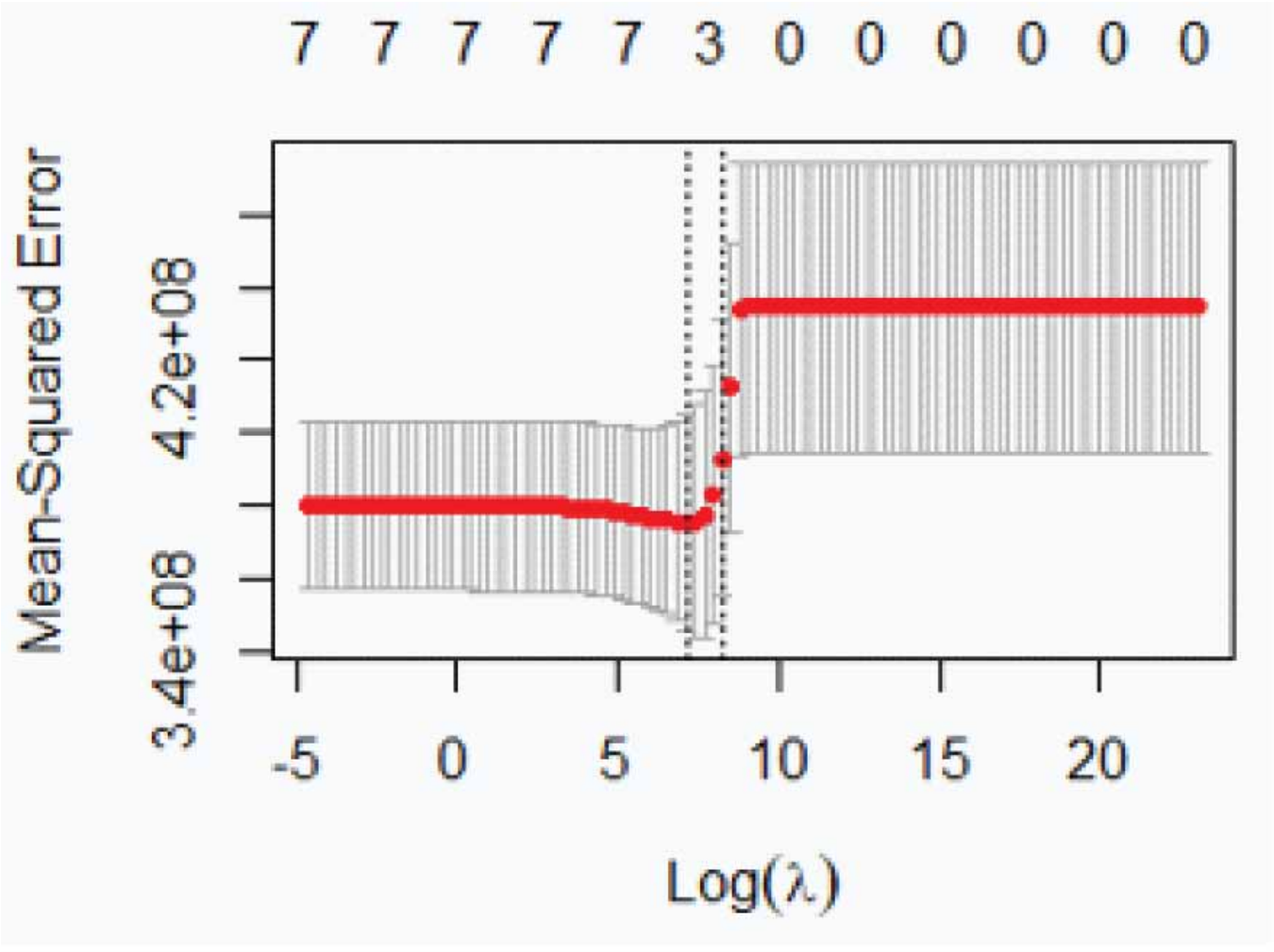

If there are n variables, then number of subsets created are:

$$2^n$$

In this method, linear regression is performed in each subset and one with best adjusted r-squared value is selected. Three types of sub-setting can be performed, i.e., forward, backward and, exhaustive. In this data backward sub-setting is implemented as it requires less computational processing. Table 10 shows the results of sub-setting performed in CDs data.

From Table 10, seventh subsets give the minimum accuracy and will help the model to tune the parameter for better optimization, and will not overfit or underfit the model.

With all these techniques, various linear regression approaches are applied and therefore, after finalizing the model, validation part would be predicted and evaluation metrics will be computed.

**Table 10.**
**Feature selection matrix**

| Subset | Feature(s) Selected | Accuracy |
|---|---|---|
| 1 | Open | 16458.03 |
| 2 | Open + MinROI | 15237.09 |
| 3 | Open + MinROI + RRR | 15251.21 |
| 4 | Open + MinROI + RRR + PRR | 14850.71 |
| 5 | Open + MinROI + RRR + PRR + CRR | 14685.41 |
| 6 | Open + MinROI + RRR + PRR + CRR + Close | 14305.72 |
| 7 | Open + MinROI + RRR + PRR + CRR + Close + MaxROI | 14194.30 |

## 5.8 Step 8: Evaluation Metric

As it is a regression problem, root mean square error (RMSE) is computed for each model built, results of each model are shown in Table 11.

Thus, these evaluation metrics concludes that sub-setting gives the best model, which can be used for prediction.

## 5.9 Step 9: Data Prediction

In this step, predicted values are interpreted to observe the difference among the target variable after building a model.

Figure 19 reveals the difference between actual volume and predicted volume, where it can be observed that in some scenario the difference is less while in some the difference is high and it depends on ROI, Sensex and bank ratios.

Thus, with these steps money market instruments are analyze with capital market and bank rates ratio to identify the difference.

**Table 11.**
**Comparison of RMSE for each model**

| Model | RMSE |
|---|---|
| Linear Regression (OLS) | 20433.17 |
| Linear Regression (Gradient Descent) | 18871.19 |
| Lasso Regression | 14888.27 |
| Sub-setting | 14194.30 |

**Figure 19.**
**Data Predicted from the model**

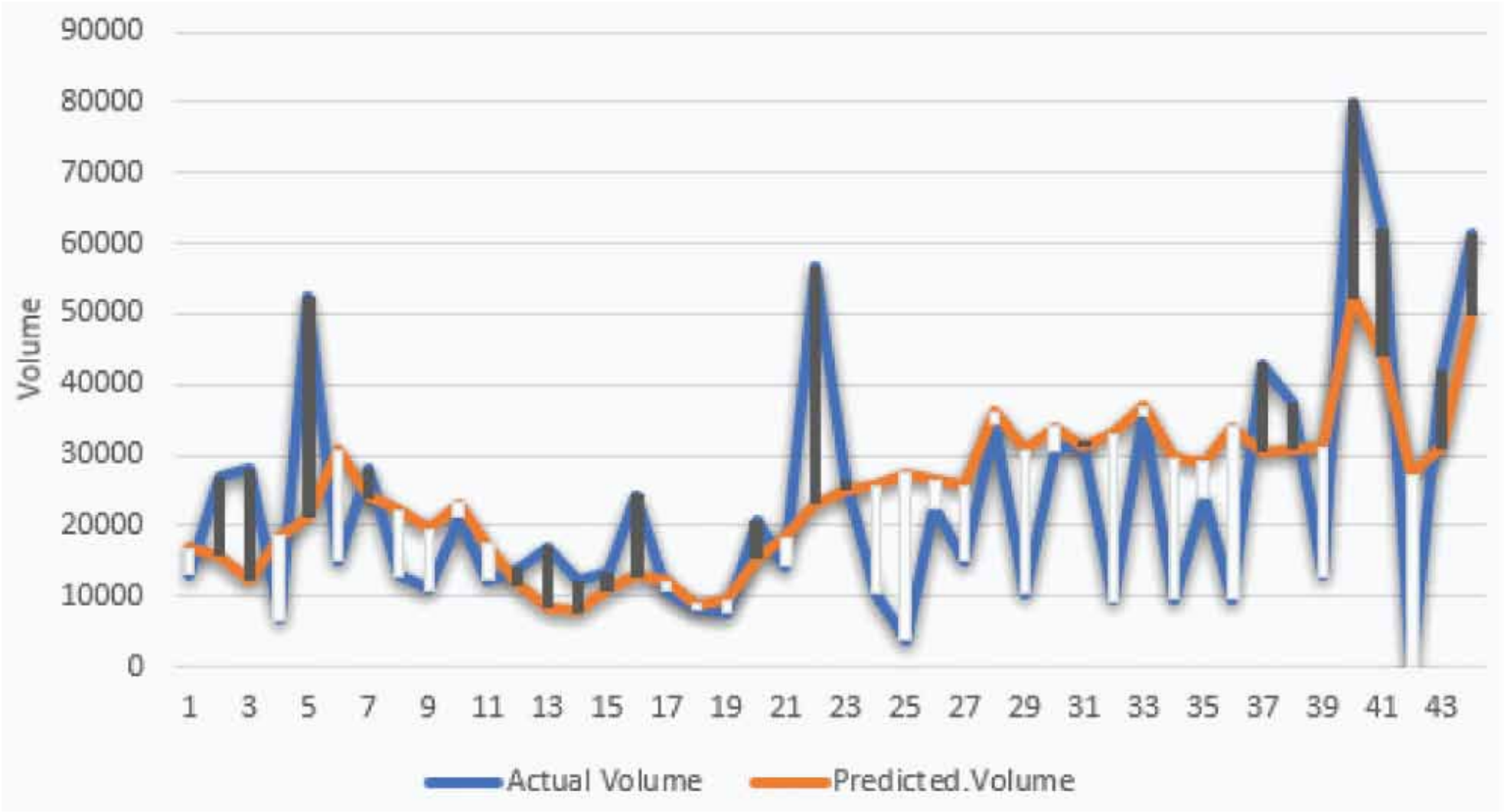

## 6. CONCLUSION AND FINDINGS

In conclusion, the deliverables which were proposed in the project are achieved successfully, summary conclusion of each deliverables are:

**Deliverable 1:** This deliverable is achieved by analyzing each variable, and creating a correlation matrix which concludes that commercial paper holds a positive correlation with stock market index and holds a negative correlation with bank rate ratios, whereas, certificate of deposits holds a negative correlation with stock market and positive correlation with bank rate ratios. Therefore, Hypothesis No. 2 is accepted that is, there is a relationship between money market with capital market and bank rate ratios.

**Deliverable 2:** This deliverable is achieved by analyzing two time series model, i.e. ARIMA model and Holt Winters model, which is used to forecast the volume of commercial papers. Also, the commercial paper is multiplicative in nature, and for the past few years the series is stationary which is proved by applying the ADF Test. Furthermore, the trend of commercial paper is increasing exponentially with some errors and monthly seasonality, and hence all the trends and anomalies are identified by decomposing the series into trend, seasonality and errors (random work).

**Deliverable 3:** This deliverable is achieved by combining all the data, building an optimized model which will predict the future price of any instruments based on past fortnight days transaction. This helps to predict the future volume based on capital market data and bank rates ratios, so that certain measures can be opted. Various techniques such as OLS, Gradient Descent, Regularization and Sub-setting method are used to build the model, so that minimal accuracy is achieved and optimized model is implemented.

In conclusion, this analysis report identifies all the trends in money market instruments, capital market and bank rates ratio, and how they are significant with each other.